\documentclass[12pt]{article}
\usepackage{paper2e}
\usepackage{mydefs2e}
\usepackage{epsf}

\begin{document}
\begin{titlepage}
\preprint{UMD-PP-00-009}

\title{Massive Higher-dimensional Gauge Fields\\\medskip
as Messengers of Supersymmetry Breaking}


\author{Z. Chacko,%
\footnote{E-mail: {\tt zchacko@bouchet.physics.umd.edu}}
\ 
Markus A. Luty,%
\footnote{Sloan Fellow.
E-mail: {\tt mluty@physics.umd.edu}}
\ 
Eduardo Pont\'on%
\footnote{E-mail: {\tt eponton@wam.umd.edu}}}

\address{Department of Physics\\
University of Maryland\\
College Park, Maryland 20742, USA}

\begin{abstract}
We consider theories with one or more compact dimensions with size
$r~>~1/M$, where $M$ is the fundamental Planck scale, with the
visible and hidden sectors localized on spatially separated
`3-branes.'
We show that a bulk $U(1)$ gauge field spontaneously broken on the
hidden-sector 3-brane is an attractive candidate for the messenger of
supersymmetry breaking.
In this scenario scalar mass-squared terms are proportional to $U(1)$
charges, and therefore naturally conserve flavor.
Arbitrary flavor violation at the Planck scale gives rise to
exponentially suppressed flavor violation at low energies.
Gaugino masses can be generated if the standard gauge fields
propagate in the bulk;
$\mu$ and $B\mu$ terms can be generated by the Giudice-Masiero
or by the VEV of a singlet in the visible sector.
The latter case naturally solves the SUSY CP problem.
Realistic phenomenology can be obtained either if all microscopic
parameters are order one in units of $M$, or if the theory is
strongly coupled at the scale $M$.
In either case, the only unexplained hierarchy is the `large' size of
the extra dimensions in fundamental units, which need only be an order
of magnitude.
All soft masses are naturally within an order of magnitude of
$m_{3/2}$, and trilinear scalar couplings are negligible.
Squark and slepton masses can naturally unify even in the absence
of grand unification.
\end{abstract}


\end{titlepage}

\section{Introduction}
The hidden-sector scenario for supersymmetry (SUSY) breaking
\cite{hidden} is arguably the simplest and most natural mechanism for
realizing SUSY in nature.
In this scenario, one assumes that there is a hidden sector in which
SUSY is broken, and that Planck-scale suppressed interactions
arising from supergravity (SUGRA) or string theory couple
the hidden and visible sectors.
Scalar masses are assumed to arise from higher-dimension operators of
the form
\beq\eql{badcont}
\De\scr{L}_{\rm eff} \sim \myint d^4\th\,
\frac{1}{M^2}(\Si^\dagger \Si) (Q^\dagger Q),
\eeq
where $Q$ is an visible sector field,
and $\Si$ is a hidden sector field with $\avg{F_\Si} \ne 0$.
Gaugino masses can arise from
\beq\eql{convgaugino}
\De\scr{L}_{\rm eff} \sim \myint d^2\th\,
\frac{1}{M} \Si W^\al W_\al + \hc,
\eeq
and $\mu$ and $B\mu$ terms can arise from the Giudice-Masiero term
\cite{GM}
\beq\eql{GM}
\De\scr{L}_{\rm eff} \sim \myint d^4\th \left[
\frac{1}{M} \Si^\dagger
+ \frac{1}{M^2} \Si^\dagger \Si \right] H_u H_d + \hc
\eeq
In addition, trilinear scalar interactions ($A$ terms) can arise from
operators of the form
\beq\eql{Aterms}
\De\scr{L}_{\rm eff} \sim \myint d^2\th\,
\frac{1}{M} \Si QQ H_{u,d} + \hc
\eeq
If $M$ is of order the Planck scale and $\avg{F_\Si} \sim (10^{11}\GeV)^2$,
this naturally generates all required soft SUSY breaking terms at the weak
scale with $m_{\rm soft} \sim m_{3/2}$.
This scenario is attractive from a theoretical point of view because
all of the ingredients are either there of necessity (\eg~supergravity) or
arise naturally (\eg~hidden sectors are a generic consequence of string
theory).
In order to obtain a gaugino mass (and $\mu$, $B\mu$ and $A$ terms) in this
way, $\Si$ must be a gauge singlet, so the hidden sector must contain a
singlet with a large $F$ term.%
\footnote{Dynamical SUSY breaking models with this feature are discussed in
\Refs{singlet}.}

The major difficulty with this scenario is that there is no
compelling reason for the interactions \Eq{badcont} that communicate
between the hidden and visible sector to preserve flavor.
Off-diagonal squark masses are severely constrained by FCNC's;
for example, mixing and CP violation in the $K$ system give%
\footnote{For a complete discussion, see \eg~\Ref{SUSYFCNC}.}
\beq
\frac{m^2_{\tilde{d}\tilde{s}}}{m^2_{\tilde{s}}}
\lsim (6 \times 10^{-3}) \left( \frac{m_{\tilde{s}}}{1\TeV} \right),
\qquad
\Im \left(\frac{m^2_{\tilde{d}\tilde{s}}}{m^2_{\tilde{s}}} \right)
\lsim (4 \times 10^{-4}) \left( \frac{m_{\tilde{s}}}{1\TeV} \right).
\eeq
An elegant solution to this problem was proposed by Randall and
Sundrum in \Ref{RS}.
They considered a situation where the hidden and visible sectors are
localized on spatially separated `3-branes'
in $D > 4$ spacetime dimensions,
with only supergravity propagating in the bulk.
(This is similar to the Ho\v rava--Witten vacuum in the
context of M theory \cite{HW}.)
\Ref{RS} pointed out that in this set-up flavor-violating interactions
between the hidden and visible sectors from short-distance physics
are suppressed even if the underlying theory does not conserve flavor.
The reason is that the exchange of particles with masses of order
the $D$-dimensional Planck scale $M_D$ (\eg~massive string states)
is exponentially suppressed by Yukawa factors $\sim e^{-M_D r}$,
where $r$ is the distance between the sectors.
A modest hierarchy $r \gsim 10 / M_D$ is therefore
sufficient to suppress flavor-changing neutral currents.

If only supergravity propagates in the bulk,
the leading contribution to soft masses is
directly related to the conformal anomaly \cite{RS,GLMR}, and
gives calculable scalar and gaugino masses proportional to
anomalous dimensions.
(It is nontrivial that exchange of supergravity KK modes
does not give rise to contact interactions of the form \Eq{badcont}.
This is discussed in detail in \Ref{LS}.)
This `anomaly mediated' scenario is attractive in that it
automatically gives flavor-diagonal scalar masses, but it
suffers from a number of drawbacks.
Most importantly, the slepton mass-squared terms are negative
in the MSSM.
Also, the Giudice-Masiero mechanism does not naturally solve
the $\mu$ problem.
There have been a number of proposals to make this scenario
realistic without spoiling its attractive features
\cite{PR,CLMP,KSS}.

In this paper, we consider a variation on this scenario that
naturally conserves flavor while preserving the
desirable features of hidden sector models described above.
Following \Ref{RS}, we consider models where SUSY is broken on a
spatially separated 3-brane.
This guarantees the absence of FCNC's from uncalculable contact
interactions of the form \Eq{badcont}.
The new ingredient we add is a $U(1)_X$ gauge multiplet that
propagates in the bulk and couples the fields in the hidden and
visible sectors.
$U(1)_X$ is assumed to be spontaneously broken by vacuum expectation
values of charged scalars localized on the hidden-sector 3-brane.
Exchange of the massive $U(1)_X$ gauge boson gives rise to
terms in the 4-dimensional effective theory of the form
\beq\eql{goodcont}
\De\scr{L}_{\rm 4,eff} \sim \myint d^4\th\, \frac{1}{v^2}
(\Si^\dagger X \Si) (Q^\dagger X Q),
\eeq
where $v$ is the VEV that breaks $U(1)_X$ and $X$ is the
charge matrix.
If $\avg{F_\Si} \ne 0$, this gives rise to visible sector
scalar mass-squared terms proportional to their $U(1)_X$ charges.
It is natural for all fields with the same standard-model 

If $U(1)_X$ commutes with flavor symmetries, then all fermions with
the same gauge quantum numbers will have the same scalar mass-squared,
which does not give FCNC's.

The SUSY breaking terms induced by $U(1)_X$ exchange preserve
$U(1)_R$, and therefore do not give rise to gaugino masses (or
$\mu$, $B\mu$, and $A$ terms).
The simplest way to generate gaugino masses is to assume the
standard-model gauge fields propagate in the bulk.
Gaugino masses can then be generated by contact terms of the form
\Eq{convgaugino} with the hidden sector brane.
More precisely, the term is the supersymmetric completion of the
operator
\beq\eql{Dgaugino}
\Delta\scr{L}_D = \de^{D - 4}(y - y_0) \Si(x)
\tr (F^{MN} F_{MN})(x, y_0) + \cdots
\eeq
where $\Si$ is the scalar component of a chiral superfield propagating
on the 3-brane at $y = y_0$, $F^{MN}$ is the field strength of
the standard-model gauge field.

There are several possibilities for generating $\mu$ and $B\mu$
terms.
One possibility is to assume that the Higgs fields
propagate in the bulk.
Then $\mu$ and $B\mu$ terms can be generated by contact terms with
the hidden sector of the form \Eq{GM};
more precisely, the supersymmetric completion of the operators
\beq\eql{Dmu}
\bal
\Delta\scr{L}_D &= \de^{D - 4}(y - y_0) \left[
(\partial^2 \Si^\dagger)(x)
+ (\partial^2 \Si^\dagger \Si)(x) \right]
(H_u H_d)(x, y_0)
\\
&\qquad + \hc + \cdots
\eal\eeq
This leads to phenomenology similar to conventional hidden sector models,
except that $A$ terms can be naturally be small because separating the hidden
and visible sectors forbids operators of the form \Eq{Aterms}.
(There are small loop-suppressed $A$ terms from anomaly-mediation
\cite{GLMR}.)

Another possibility is that there is a singlet field $S$ whose VEV
generates effective $\mu$ and $B\mu$ terms \cite{NMSSM}.
We assume that the 4-dimensional effective theory includes the cubic
superpotential terms
\beq
\De W = \la S H_u H_d + \frac{\ka}{3} S^3.
\eeq
The $S^3$ term is not $U(1)_X$ invariant, but non-invariant terms such as
this can be present below the scale where $U(1)_X$ is broken.
(We will see that this requires the Higgs fields
to propagate in the bulk, while $S$ can propagate either in the bulk
or on the hidden-sector brane.)
The fields $H_{u,d}$ and $S$ can naturally have a negative mass-squared of
order the weak scale, giving rise to realistic electroweak symmetry breaking.
An attractive feature of this model in the present higher-dimensional
context is that it automatically solves the SUSY CP problem \cite{SUSYCP}:
all CP-violating phases can be rotated into the CKM phase and
$\vartheta_{\rm QCD}$.
For this it is crucial that there are no uncontrolled $A$ terms from
higher-dimension operators.

The couplings of the bulk fields such as the gauge fields (and
possibly Higgs fields) in the effective 4-dimensional theory will
be suppressed by the volume of the compact subspace, so $r$
cannot be too large.
On the other hand, we have seen above that $r$ must be sufficiently
large so that FCNC's are suppressed.
Since the suppression of FCNC's is exponential, these requirements
are easily met, especially for a small number of large extra dimensions.
We will show below that even for a large number of extra dimensions
(\eg~$D = 11$) these requirements can be met if the
standard-model gauge couplings are strongly coupled at the fundamental
Planck scale.

In order that the visible sector scalar masses be close to the
gaugino masses, the VEV $v$ that breaks $U(1)_X$ must also be
near the fundamental Planck scale.
This emerges naturally if all couplings are order 1 in units of the
fundamental Planck scale.
(In fact, the bulk standard-model gauge couplings must be somewhat
larger than this in order for the effective 4-dimensional gauge
couplings to be order 1, but this factor need not be larger than
an order of magnitude.)

Alternatively, we can consider a scenario where \emph{all}
microscopic couplings are near their strong-coupling values at
the fundamental scale.
In this scenario, the only large parameter is the size of the compact
dimensions, which need only be an order of magnitude larger than the
fundamental scale.
We carry out estimates of the size of parameters in this scenario,
paying attention to geometrical factors that control the size of
loop graphs (`\naive dimensional analysis' \cite{NDA,SUSYNDA}).
The result is that this scenario naturally gives scalar masses,
gaugino masses, and $\mu$ and $B\mu$ terms of realistic size without
the introduction of small parameters.

This strongly-coupled scenario is particularly appealing in the
light of the modern view of string theory as a single connected moduli
space of different theories, with the known 10-dimensional superstring
theories and 11-dimensional SUGRA appearing as weak coupling limits
\cite{Mtheory}.
Already in the early days of string theory it was realized that it
is extremely difficult to find phenomenologically viable vacua
near weak coupling because the theory generally runs away to zero
coupling \cite{weak}.
With the modern picture in mind, one can conjecture that
phenomenologically viable vacua exist in the regime
where the theory has no weak-coupling description.
But then the absence of FCNC's appears especially puzzling, since
we expect all operators allowed by gauge symmetries to be generated
with approximately equal strength.
The present class of models gives a possible solution:
if the vacuum contains 3-branes, and some compact dimensions are
an order of magnitude larger than the string scale, this can act as
a `seed' for accidental symmetries in the low-energy world.
There are other small parameters that are not directly explained in
this approach (such as the small Yukawa couplings);
it would be interesting to see whether these small parameters
can also have a geometric origin in a scenario of this type.

This paper is organized as follows.
In Section 2, we discuss the physics of breaking gauge
symmetry on branes.
The considerations are elementary, but there are some surprises.
In Section 3, we write down explicit models and give estimates for
soft masses.
Section 4 contains our conclusions.

\section{Breaking Gauge Symmetry on Branes}
In this Section, we discuss the breaking of a bulk gauge symmetry
by the VEV's of charged fields propagating on 3-branes.
This is simpler than breaking the gauge symmetry in the bulk because
the allowed interactions of supersymmetric theories
in higher dimensions are quite limited.

We will be interested in operators in
the 4-dimensional low-energy theory of the form
\beq
\De\scr{L}_4 \sim \myint d^4\th\, \Si^\dagger \Si \left[
Q^\dagger Q + H^\dagger H \right],
\eeq
where $\Si$ is a field propagating on the hidden sector brane,
$Q$ is a field propagating on the visible sector brane, and
$H$ is a bulk field.
To compute the coefficient of quartic terms such as these, it suffices
to compute the 4-fermion component.
Below we will compute the contribution to the 4-fermion term
from the tree-level exchange of vector bosons, where the couplings
are completely determined by gauge invariance.
In supersymmetric theories in more than 4 dimensions, there are
additional propagating bosonic fields in the gauge multiplet that
could in principle contribute to the coefficient of the 4-fermion
term, but we will later give an explicit example where we show that
only the vector boson contributes.
We believe that this feature is more general, but establishing this
would require a general understanding of the couplings between
higher-dimensional supersymmetric gauge fields to branes.
We will not address this question here.

We consider two parallel 3-branes in a $D$-dimensional space,
with $D - 4$ dimensions compactified on a length scale $r$.
We assume that there is a $U(1)_X$ gauge field in the bulk, and 
charged fermions on the branes and in the bulk.%
\footnote{The conclusions of this Section hold for non-Abelian
gauge theories as well.}
The lagrangian is
\beq\eql{nosusyL}
\bal
\!\!\!\!\!\!
\scr{L}_D &= -\frac{1}{4 g_D^2} (F^{MN} F_{MN})(x, y)
+ \bar{\Psi}_B i\Ga^M D_M \Psi_B
\\
&\qquad
+ \de^{D - 4}(y - y_1) \left[
(D^\mu \phi^\dagger D_\mu \phi)(x) - V(\phi(x))
+ (\bar{\psi}_1 i\,\ga^\mu D_\mu \psi_1)(x) + \cdots \right]
\\
&\qquad
+ \de^{D - 4}(y - y_2) \left[
(\bar{\psi}_2 i\, \ga^\mu D_\mu \psi_2)(x) + \cdots \right].
\eal\eeq
Here $M, N, \ldots = 0, \ldots, D - 1$ are Lorentz indices for the
$D$-dimensional spacetime, $x^\mu$ ($\mu = 0, \ldots, 3$) are coordinates
along the 3-brane, and $y^I$ ($I = 4, \ldots, D - 1$) are coordinates
for the compact space.
We will assume that $\avg{\phi} \ne 0$, and we will work out the
interactions between the fermions induced by $U(1)_X$ gauge exchange.
For the bulk fermion, we are interested only in the zero mode
\beq
\Psi_B(x, y) = \frac{1}{\sqrt{V_{D - 4}}} \psi_B(x) + \cdots,
\eeq
where $V_D$ is the volume of the compact space.
The gauge fields $A_M$ are normalized to have mass dimension $+1$,
so that covariant derivatives are given by
\beq
D_M = \partial_M - i A_M X,
\eeq
where $X$ is the $U(1)_X$ charge matrix.
The gauge coupling $g_D$ has mass dimension $(4 - D)/2$.

The lagrangian \eq{nosusyL} makes sense as an effective lagrangian
valid below an ultraviolet cutoff $\La_0$ provided that we include
all counterterms allowed by the symmetries proportional to powers
of $\La_0$.
In particular, the lagrangian will include terms of the form
\beq\eql{ct}
\bal
\!\!\!\!\!\!\!\!
\De\scr{L}_D \sim & \frac{(\bar{\Psi}_B \Psi_B)^2}{\La_0^{D - 2}}
+ \de^{D - 4}(y - y_1) \left[
\frac{ (\bar{\psi}_1 \psi_1)^2}{\La_0^2}
+ \frac{(\bar{\psi}_1 \psi_1) (\bar{\Psi}_B \Psi_B)}{\La_0^{D - 2}}
+ \frac{ (\bar{\Psi}_B \Psi_B)^2}{\La_0^{2(D - 3)}}  \right]
\\
&
+\, (1 \leftrightarrow 2),
\eal
\eeq
so that contact terms of this form are not calculable unless their
coefficients are parametrically larger than those above.
Note that the $D$-dimensional theory cannot have contact term of the
form $(\bar{\psi}_1 \psi_1) (\bar{\psi}_2 \psi_2)$ by locality, so
interactions of this form are always calculable.

We assume that the scalar field on the brane at $y = y_1$ gets a
VEV that breaks the gauge symmetry:
\beq
\avg{\phi} = \frac{v}{\sqrt{2}}.
\eeq
It will be convenient for us to normalize the $U(1)_X$ charges by
taking the scalar field $\phi$ (and hence $v$) to have charge
$+1$.

To fix the gauge, we add a bulk gauge-fixing term proportional to
$(\partial^M A_M)^2$ to make the gauge kinetic term canonical:
\beq
\scr{L}_D = -\frac{1}{2 g_D^2} \partial^M A^N \partial_M A_N + \cdots.
\eeq
This does not completely fix the gauge, since we can still make
gauge transformations with gauge parameter $\al$ satisfying
$\partial^M \partial_M \al = 0$.
We can use this residual gauge freedom to choose unitary gauge for
the scalar field on the brane at $y = y_1$.
In this gauge, we can expand the gauge fields in Kaluza-Klein (KK)
modes:
\beq
A^\mu(x, y) = \sum_k A^\mu_k(x) f_k(y),
\qquad
A^I(x, y) = \sum_k A^I_k(x) g_k(y),
\eeq
where the KK wavefunctions satisfy
\beq\bal
-\nabla_{\!\perp}^2 f_k(y) + g_D^2 v^2 \de^{D - 4}(y - y_1) f_k(y)
&= m^2_k f_k(y),
\\
-\nabla_{\!\perp}^2 g_k(y) &= \tilde{m}^2_k g_k(y),
\eal\eeq
where $\nabla_{\!\perp}^2$ is the Laplacian in the compact space,
and $m_k$, $\tilde{m}_k$ are the masses of the KK modes.

Note that the VEV on the 3-brane does not affect the $A^I$ fields,
so there are in general $D - 4$ massless scalars in the 4-dimensional
theory.
In a supersymmetric theory, there are additional fields in the
bulk gauge multiplet (\eg~gauginos) that do not acquire a mass from
the VEV on the 3-brane.
This can be avoided by an orbifold projection that gives mass to these
fields, as in the specific model discussed in Section 3.1.
More generally, some of these states may survive to the weak scale
and have phenomenological consequences.
This will be discussed in Section 3.4.

The effect of the VEV on the fields $A^\mu$ depends on the spacetime
dimension $D$ as well as the choice of the parameters.
The dimensionless measure of the relevance of
the delta function term as a perturbation is
\beq
\ep \equiv \frac{g_D^2 v^2 r^2}{V_{D - 4}} \sim g_4^2 v^2 r^2.
\eeq
We will always be interested in $r \gg 1/v$ (`large' extra dimensions),
so we see that $\ep \gg 1$ if $g_D$ is large enough so that $g_4 \sim 1$.
(We will later show that this is possible even for large $D$.)
However $\ep \ll 1$ if $g_D^2 \sim 1/v^{D - 4}$ for $D \ge 6$.
It is therefore natural to consider both large and small $\ep$ for
arbitrary dimensions. 

If $\ep \ll 1$, we can use perturbation theory to find the mass
of the lightest KK mode.
The unperturbed KK wavefunction is simply a constant zero mode,
and we find
\beq
m^2_0 = \frac{g_D^2 v^2}{V_{D - 4}} = g_4^2 v^2,
\eeq
where $g_4$ is the gauge coupling in the 4-dimensional effective theory.
This is the result we would expect from the Higgs mechanism in the
4-dimensional low-energy theory.
Intuitively, the picture is that the zero mode of the gauge field
is constant across the extra dimension, and therefore feels the
VEV on the other wall as if it were 4-dimensional.

\begin{figure}[t]
\centering
\centerline{\epsfxsize=5.0in\epsfbox{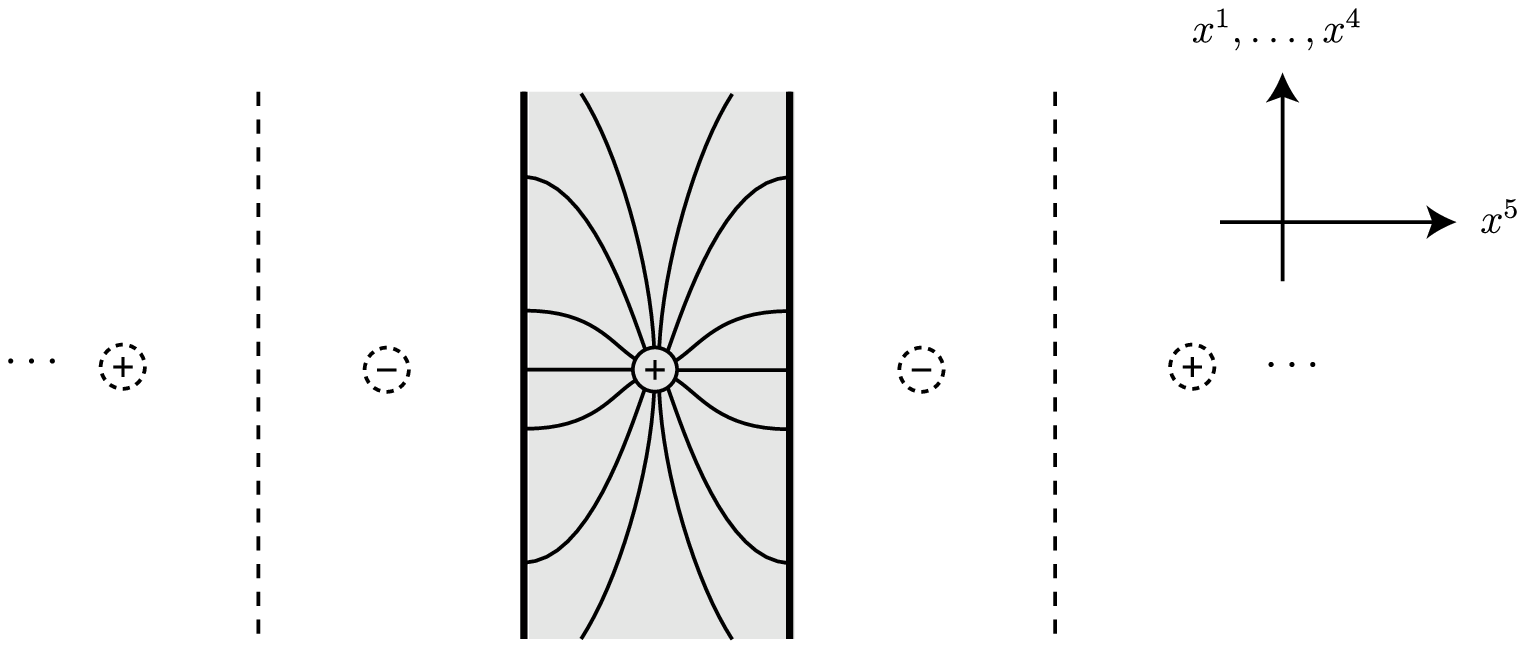}}
\smallskip
\capt{Electric field due to a point charge in a space with one
compact dimension.
The two solid vertical lines represent two copies of the 3-brane
where $U(1)_X$ is broken;
the physical space is the shaded region between them.
The field can be reproduced by an infinite number of image charges
of alternating signs, as shown.}
\end{figure}

For $\ep \gg 1$, we cannot treat the delta function as a
perturbation and the situation is quite different.
We can understand this limit intuitively by noting that the 3-brane on
which the $U(1)_X$ is broken is a superconductor.
We can compute the mass of the lightest KK mode from the fact that it
controls the exponential fall-off of the electric field due to a point
charge at distances large compared to $r$.
We expect a non-zero cutoff-independent mass, since the result is already
nonzero and cutoff-independent for $\ep \ll 1$.
This is illustrated in Fig.~1 for the case $D = 5$.
The field will be distorted by induced surface charges on the superconducting
3-brane that tend to screen the electric field.
In the limit $\ep \to \infty$ the brane acts as a perfect conductor, and
the problem of finding the electric field of a point
charge is purely geometric, with the superconducting 3-brane acting
as a boundary condition.
Just by dimensional analysis the only possible result is
\beq
m^2_0 \sim \frac{1}{r^2}.
\eeq
In fact, for $D = 5$ the electrostatics problem described here
can be solved by the method of images (see Fig.~1).
The infinite number of image charges is responsible for the exponential
fall-off of the electric field.
For finite $\ep$, we expect corrections to this picture suppressed
by inverse powers of $\ep$.

The qualitative considerations above are born out
by direct calculation of the KK decomposition.
For example, for $D = 5$ with the extra dimension compactified on
a circle of radius $r$, we have
\beq\eql{D5KK}
f_k(y) = \sin\left(\frac{k y}{r}\right) + \frac{2k}{g_5^2 v^2 r}
\cos\left( \frac{k y}{r} \right),
\qquad
y \le 0 < 2\pi r,
\eeq
where
\beq
m^2_k = \frac{k^2}{r^2}
\eeq
and the allowed values of $k$ are determined by
\beq
k \tan(\pi k) = \sfrac{1}{2} g_5^2 v^2 r.
\eeq
For $\ep = g_5^2 v^2 r \gg 1$, the solutions are
\beq
k = (n + \sfrac{1}{2}) \left[ 1 + O(\ep^{-1}) \right],
\quad
n = 0, 1, \ldots
\eeq
so the masses of the lowest-lying KK modes are of order $1/r$ as
claimed.

For large values of $D$ the KK decomposition is more complicated,
and it is simpler to compute the contact terms we are interested
in directly from the higher-dimensional theory.
We perform the tree-level calculation by solving the equations of motion
in the compact directions
\beq
\!\!\!\!\!\!
-\frac{1}{g_D^2} \nabla_{\!\perp}^2 A^\mu
+ \de^{D - 4}(y - y_1) \left[ v^2 A^\mu + J_1^\mu \right]
+ \de^{D - 4}(y - y_2) J_2^\mu + \frac{J_B^\mu}{V_{D - 4}} = 0,
\eeq
where
\beq
J_{1,2}^\mu = \bar{\psi}_{1,2} \ga^\mu X \psi_{1,2},
\qquad
J_B^\mu = \bar{\psi}_{B} \ga^\mu X \psi_{B},
\eeq
are treated as background fields.
We are mainly interested in the coefficient of the $J_1 J_2$ term,
and we claim that for any dimension and for any value of $\ep$, 
this coefficient is of order $1/v^2$.
If $\ep \ll 1$, this is not surprising since the lightest KK mode
has mass of order $g_4^2 v^2$, and exchange of this mode
gives rise to a contact interaction with strength
$g_4^2 / (g_4^2 v^2) \sim 1/v^2$.
For $\ep \gg 1$ this result is perhaps counterintuitive since one might
expect the coefficient to be
$g_4^2 / m_{\rm KK}^2 \sim g_D^2 / r^{D - 6}$.

We begin with $D = 5$ compactified on a circle of radius $r$,
with the branes at $y = 0$ and $y = \pi r$.
The solution is
\beq
A^\mu(y) =
\begin{cases}
\displaystyle
\al_1^\mu + \be_1 ^\mu y + \frac{1}{4\pi r} g_5^2 J_B^\mu y^2
& $0 \le y < \pi r$,
\\
\displaystyle
\al_2^\mu + \be_2^\mu y + \frac{1}{4\pi r} g_5^2 J_B^\mu y^2
& $\pi r \le y < 2\pi r$,
\end{cases}
\eeq
where the coefficients $\al_{1,2}$ and $\be_{1,2}$ are determined
by matching the value of $A^\mu$ and the discontinuity in its
derivative at the positions of the delta functions.
The result of solving the equations and substituting back into
the lagrangian is
\beq\eql{JJ5}
\De\scr{L}_{4,{\rm eff}} = -\frac{1}{2v^2} (J_1 + J_2 + J_B)^2
- \frac{\pi g_5^2 r}{4} \left(
J_2^2 + J_2 J_B + \sfrac{1}{3} J_B^2 \right).
\eeq
Note that the coefficient of the $J_1 J_2$ and $J_1 J_B$ terms is
of order $1/v^2$ rather than
$g_4^2 / m_{\rm KK}^2 \sim g_D^2 / r^{D - 6} \gg 1/v^2$, as
might be expected.
This is because the overlap of the low-lying KK modes with the brane
where $U(1)_X$ is broken is suppressed.
\Eq{D5KK} shows that the this overlap is suppressed
by $1/\ep = 1/(g_5^2 v^2 r)$ and the contribution of a low-lying KK mode to
the $J_1 J_2$ or $J_1 J_B$ term is of order
\beq
\frac{g_4^2 / \ep}{m_{\rm KK}^2} \sim \frac{1}{v^2}.
\eeq

We next consider $D = 6$ compactified on a sphere with the branes at
opposite poles.
The classical solution for $A^\mu$ then depends only on the
azimuthal angle $\th$.
To make the equations well-defined, the delta functions must be
regulated.
We do this by replacing the point-like delta functions by a delta
function ring of finite radius.
Specifically, in terms of the variable $z \equiv \cos\th$ we have
\beq
\de^2(y - y_{1,2}) \to \frac{1}{2\pi r^2} \de(z - z_\pm),
\qquad
z_\pm \equiv \pm(1 - \de),
\eeq
where $\de > 0$ is a small cutoff parameter.
The equations of motion are
\beq\bal
-\frac{1}{g_6^2} \frac{d}{dz} \left[
(1 - z^2) \frac{d A^\mu}{dz} \right]
& + \frac{1}{2\pi} \de(z - z_+) \left[
v^2 A^\mu + J_1^\mu \right]
\\
& + \frac{1}{2\pi} \de(z - z_-) J_2^\mu
+ \frac{1}{4\pi} J_B^\mu = 0.
\eal\eeq
Note that $r$ factors out;
this system is classically scale invariant.
The solution is
\beq
A^\mu(z) =
\begin{cases}
\displaystyle
\al_1^\mu - \frac{g_6^2}{4\pi} J_B^\mu \ln(1 - z)
& $-1 \le z < z_-$,
\\
\displaystyle
\al^\mu + \be^\mu \ln\frac{1 + z}{1 - z}
- \frac{g_6^2}{8\pi} J_B^\mu \ln(1 - z^2)
& $z_- \le z \le z_+$,
\\
\displaystyle
\al_2^\mu - \frac{g_6^2}{4\pi} J_B^\mu \ln(1 + z)
& $z_+ < z \le 1$.
\end{cases}
\eeq
As before, the coefficients $\al_{1,2}$ and $\al, \be$ are
determined by matching the value of $A^\mu$ and the discontinuity
in its derivative at the delta functions.
Substituting back into the lagrangian, we obtain
\beq\eql{JJ6}
\bal
\De{\scr L}_{\rm eff, 4} = -\frac{1}{2 v^2} (J_1 &+ J_2 + J_B)^2
+ \left( \frac{g_6^2}{4\pi} \ln\frac{\de}{2} \right) (J_2^2 + J_2 J_B)
\\
& + \frac{g_6^2}{8 \pi} \left( 1 + \ln\frac{\de}{2} \right) J_B^2
+ O(\de).
\eal\eeq
The physical origin of the logarithmic divergences is that the brane
acts as a source for modes with wavelength of order the brane thickness.
The logarithmic divergences can be absorbed into counterterms of the form
\Eq{ct}.
Note that, as in the $D = 5$ case, all contact terms involving
$J_1$ are of order $1/v^2$ for arbitrary $\ep$.
Although the spherical geometry is clearly a special case, we
believe that the qualitative features are quite general.%
\footnote{To be realistic, a spherical geometry would require a
source for the curvature of spacetime.
In a supersymmetric theory, such a source of curvature would in
general break SUSY.
We do not enter into these considerations, since we are using the
spherical geometry only for illustrative purposes.}

It is easy to generalize this calculation to higher dimensions 
in the case where the compact space
is a $(D - 4)$-sphere of radius $r$ and the positions of the
3-branes are at opposite poles.
We do not include a charged bulk fermion for simplicity.
As before we use the variable $z = \cos\th$ where $\th$ is the
azimuthal angle, and regulate the delta functions by replacing them
with delta function rings at $z_\pm = \pm(1 - \de)$.
The equations of motion are
\beq\bal
-\frac{1}{(1 - z^2)^{(D - 6)/2}} &\frac{\partial}{\partial z} \left[
(1 - z^2)^{(D - 4)/2} \frac{\partial A_\mu}{\partial z} \right]
\\
& + \frac{g_D^2}{r^{D - 6}}\,
\frac{\de(z - z_+)}{\Om_{D - 4} (1 - z_+^2)^{(D - 6)/2}}
\left[ v^2 A_\mu + J_{1\mu} \right]
\\
& + \frac{g_D^2}{r^{D - 6}}\,
\frac{\de(z - z_-)}{\Om_{D - 4} (1 - z_-^2)^{(D - 6)/2}}
J_{2\mu} = 0,
\eal\eeq
where $\Om_n = 2 \pi^{n/2} / \Gamma(n/2)$ is the volume of the
unit $(n - 1)$-sphere.
The solution is
\beq
{A_\mu}(y) =
\begin{cases}
\al_{1\mu} & $-1 \le z < z_-$, \\
\al_\mu + \be_\mu f_D(z) & $z_- \le z \le z_+$, \\
\al_{2\mu} & $z_+ < z \le 1$,
\end{cases}
\eeq
where
\beq
f_D(z) = \int^z dz\, \frac{1}{(1 - z^2)^{(D - 4)/2}}.
\eeq
The coefficients $\al_{1,2}$ and $\al, \be$ can be determined by matching
the value of $A^\mu$ and the discontinuity in its derivative
at the delta function shells.
Substituting back into the lagrangian, we obtain
\beq\eql{JJD}
\De{\scr L}_{\rm eff, 4} = -\frac{1}{2 v^2} (J_1 + J_2)^2
- \frac{g_D^2 f_D(z_-)}{2 \Om_{D - 4} r^{(D - 6)/2}} J_2^2.
\eeq
The coefficient of $J_2 J_2$ diverges in the limit $\de \to 0$.
In terms of a physical length cutoff (brane thickness) $a$,
we have $\de \sim (a/r)^{1/2}$ and hence
$f_D(z_-) \sim (r/a)^{(D - 6)/2}$.
The coefficient of the $J_2^2$ term is therefore of order
$g_D^2 / a^{(D - 6)/2}$.
This divergence can be absorbed by a counterterm of the form \Eq{ct}.

We have been considering contact terms between fields on the
brane where $U(1)_X$ is broken and a spatially separated brane.
However, we could consider contact terms between fields on
spatially separated branes, neither of which is the one on which
the $U(1)_X$ is broken.
In that case, there is no suppression of the KK wavefunctions and
the coefficient of $J_2 J_3$ is of order $g_4^2 / m_{\rm KK}^2$,
as expected.

\section{Realistic Models}
We now apply the results above to the construction of realistic models.

\subsection{An Explicit Model}
We begin by giving an explicit model in 5 spacetime dimensions as
an existence proof.
We believe that these ideas can be made to work in more general
settings (\eg~in models with additional `large' dimensions).

We assume that the 5-dimensional theory has minimal supersymmetry,
namely 8 real supercharges.
One dimension is compactified on a $Z_2$ orbifold with the hidden
and visible sector fields localized on the orbifold fixed points.
The orbifold projection explicitly breaks half the supersymmetry,
which gives unbroken $\scr{N} = 1$ SUSY in the 4-dimensional
effective theory.
This makes it simple to construct couplings to the
orbifold boundary, since these need only preserve the unbroken
$\scr{N} = 1$ supersymmetry.
This set-up was analyzed in \Ref{MP}, and we make use of the
formalism described in that paper.
This scenario is also closely related to the one advocated by
Ho\v rava and Witten in the context of M theory \cite{HW}.

There are two types of 5-dimensional multiplets used to construct
the model, and we now describe them briefly.
We use the conventions of \Ref{MP}, which should be consulted
for more detail.
A gauge multiplet $(\Phi, A_M, \la^j, X^a)$ consists of a real
scalar $\Phi$, a gauge field $A_M$, a symplectic Majorana spinor
$\la^j$ ($j = 1,2$), and real auxiliary fields $X^a$ ($a = 1,2,3$).
The indices $j$ and $a$ are doublet and triplet indices for an
$SU(2)_R$ symmetry.
The fields that are even under the orbifold parity form
an $\scr{N} = 1$ gauge multiplet $(A_\mu, \la_L^1, D)$, where
\beq
D \equiv X^3 - \partial_5 \Phi
\eeq
is the auxiliary field.
(Note that this formalism forces us to use Wess--Zumino gauge for
the induced $\scr{N} = 1$ gauge multiplet.)

A 5-dimensional hypermultiplet $(\phi^j, \psi, Y^j)$ consists of 2
complex scalars $\phi^j$, a Dirac spinor $\psi$, and 2 complex
auxiliary fields $Y^j$.
The fields that are even under the orbifold parity form
an $\scr{N} = 1$ chiral multiplet $(\phi^1, \psi_L, F)$, where
\beq
F \equiv Y^1 - \partial_5 \phi^2
\eeq
is the auxiliary field.

These results make it simple to couple even-parity bulk fields to
4-dimensional fields propagating on the boundary.
For example, the coupling of the bulk $U(1)_X$ gauge field to charged
fields on the boundary can be written
\beq
\De\scr{L}_5 &= \de(y - y_1) \myint d^4\th\,
Q^\dagger e^{V X} Q
\nonumber\\
\eql{gaugewall}
&= \de(y - y_1) \left[ (D^\mu \tilde{Q}^\dagger D_\mu \tilde{Q})
+ Q^\dagger i\si^\mu D_\mu Q
+ (X^3 - \partial_5 \Phi) \tilde{Q}^\dagger X \tilde{Q} \right].
\eeq
where $V$ denotes the even-parity $\scr{N} = 1$ gauge multiplet
obtained from the bulk $U(1)_X$ multiplet, evaluated at $y = y_1$.

The couplings above are to be used to compute the contributions from
exchange of massive $U(1)_X$ gauge fields between the orbifold
boundaries.
From \Eq{gaugewall} we see that only $A_\mu$ couples to the fermions,
so the calculations of Section 2 give the correct coefficient of
the resulting 4-fermion terms.

Note that the vector polarization $A_5$ and the `extra' gaugino $\la^2$
have masses of order $1/r$ by the orbifold projection.
The $\scr{N} = 1$ gaugino $\la^1$ gets a mass from the supersymmetric
Higgs mechanism (since SUSY is not broken by the boundary VEV).
Therefore, there are no extra light states in this model.

We now consider the higher-dimension operators that couple the
bulk and boundary fields.
We will consider the case where the Higgs fields propagate in the
bulk in order to illustrate the required couplings.
We assume that each Higgs multiplet arises from a separate bulk
hypermultiplet.
We are interested in the higher-dimension couplings
\beq\eql{Cs}
\bal
\De\scr{L}_5 \sim \de(y - y_2) & \left\{
\myint d^4\th\, \left[
\Si^\dagger H_u H_d
+ \Si^\dagger \Si ( H_u H_d
+ H_u^\dagger H^{\vphantom\dagger}_u + H_d^\dagger H^{\vphantom\dagger}_d) 
\right] \right.
\\
&\quad\left.
+\, \myint d^2\th\, \Si W^\al W_\al + \hc \right\}
\eal\eeq
where $H_{u,d}$ are $\scr{N} = 1$ chiral multiplets arising from
the even-parity components of the bulk Higgs fields, and $W_\al$ is
the $\scr{N} = 1$ standard-model gauge field strength arising from the
even-parity components of the bulk gauge multiplet.
Upon matching to the effective 4-dimensional theory,
these operators give rise to effective operators of the form
\Eqs{convgaugino} and \eq{GM}.
These operators give rise to gaugino masses and $\mu$ and $B\mu$
terms when SUSY is broken by $F_\Si \ne 0$.

We now turn to squark and slepton masses.
If $U(1)_X$ is broken in the visible sector, there are flavor-violating
operators of the form
\beq
\myint d^4\th\, \frac{c_{jk}}{M^{2n}} (\Phi^\dagger e^{V X_\Phi} \Phi)^n
Q_j^\dagger e^{V X_Q} Q_k,
\eeq
where $\Phi$ is the field whose VEV breaks $U(1)_X$.
Since (as we will see) $\avg{\Phi} = v \sim M$, these operators are
unsuppressed at low energies.
There is no reason for these operators to conserve flavor, so these
operators will give rise to generation-dependent couplings of the $U(1)_X$
boson.%
\footnote{We thank R. Rattazzi for pointing this out to us.}
We must therefore break $U(1)_X$ on the hidden sector brane.
In that case, squark and slepton masses are generated by the operator
\beq
\De\scr{L}_4 \sim \frac{1}{v^2} \myint d^4\th\,
(\Si^\dagger X \Si) (Q^\dagger X Q)
\eeq
generated by $U(1)_X$ exchange.
This term will conserve flavor if $U(1)_X$ commutes with the
flavor symmetry.

Since we want all the squark and slepton masses to be positive,
the signs of the $U(1)_X$ charges of all squark and sleptons must
be the same.
This means that the $U(1)_X$ gauge field necessarily has mixed anomalies
with the standard-model gauge group.
This is not inconsistent because the $U(1)_X$ gauge group is broken at
the scale $v \sim M$, so the consistency of the low-energy field theory
is really all that is required if one is willing to put off the
derivation of the model from string theory.
However, it is reassuring to note that there is no difficulty in
constructing field theories above the scale $v$ that are free
from gauge anomalies.
Anomalies of the type $U(1)_X SU(3)_C^2$, $U(1)_X SU(2)_W^2$, and
$U(1)_X U(1)_Y^2$ can be canceled by adding chiral fields that are
in vector-like representations of the standard-model gauge group,
but chiral with respect to $U(1)_X$.
Anomalies of the type $U(1)_X^2 U(1)_Y$ can be simply canceled if
all fields charged under the standard-model have the same value of
the $U(1)_X$ charge, in which case the cancellation of these anomalies
follows from the relation $\tr(Y) = 0$.
All the extra fields added in this way can obtain $U(1)_X$-violating
masses at the scale $v$.%
\footnote{Since $U(1)_X$ and the standard-model gauge groups propagate
in the bulk, the fields that cancel the anomalies can be localized on
a distance wall.}

We now consider the $U(1)_X$ invariance of the visible sector Yukawa terms.
Because $U(1)_X$ is broken in the hidden sector, the simplest possibility
is that the Yukawa terms are invariant under $U(1)_X$.
For the small Yukawa couplings one can contemplate the possibility
that the Yukawa couplings arise from small $U(1)_X$ breaking effects
(such as the VEV of a `flavon' field charged under $U(1)_X$).
However, it seems implausible that the order-1 top Yukawa coupling
arises in this way.
If we assume that the top Yukawa coupling is $U(1)_X$ invariant, then
there is a contribution to the up-type Higgs scalar mass-squared
\beq
\De m^2_{Hu} = -(m^2_{\tilde{t}L} + m^2_{\tilde{t}R}).
\eeq
If the other Yukawa couplings also arise from $U(1)_X$ invariant
effects we have
\beq
\De m^2_{Hu} = -(m^2_{\tilde{Q}L} + m^2_{\tilde{u}R}),
\quad
\De m^2_{Hd} = -(m^2_{\tilde{Q}L} + m^2_{\tilde{d}R}).
\eeq
(The squark and slepton masses of different generations are universal.)
Given experimental bounds on squark masses, this may require moderate
fine-tuning of other contributions to the Higgs masses (\eg~from
the $\mu$ term) to obtain realistic electroweak symmetry breaking.


Our next task is to estimate the size of soft SUSY breaking in this
model.
To do this we will need to know how to estimate parameters in
strongly-coupled theories in higher dimensions, and we address this
question next.

\subsection{Na\"\i ve Dimensional Analysis in Higher Dimensions}
We argued in the Introduction that it is an attractive possibility
that all the couplings of the theory are strong at a fundamental
scale $\La$, which may be identified with the fundamental scale
in strongly-coupled M theory.
Apart from this, it is important to estimate the maximum possible
value of the $D$-dimensional gauge coupling, since this determines the
maximum size of the extra dimensions.
With this motivation, we explain how to estimate the size of
terms in the effective theory under the assumption that the
fundamental theory is strongly coupled and contains no small
parameters, generalizing previous results \cite{NDA,SUSYNDA} to
theories in higher dimensions with branes.
In such theories, one might expect that all couplings in the effective
theory below the scale $\La$ are of order 1 in units of $\La$.
However, experience with QCD and exactly solvable supersymmetric
models \cite{SWNDA} shows that there are large hierarchies in the
effective couplings when they are expressed in units of the scale
where the theory is strongly coupled.
As we will explain, these can be understood from the condition that
all interactions in the effective theory get strong at the same
scale.
These factors are related to the phase-space factors in loop integrals,
and are therefore strongly dimension-dependent, so the generalization
is non-trivial.

In a $D$-dimensional theory, a typical loop integral can be written
\beq
\myint \frac{d^D\! P}{(2\pi)^D}\, f(P^2)
\sim \frac{\Om_D}{2(2\pi)^D} \myint  dP^2\, P^{D - 2} f(P^2).
\eeq
This means that every loop factor is kinematically suppressed by
a factor of order
\beq
\frac{1}{\ell_D} = \frac{1}{2^D \pi^{D/2} \Ga(D/2)}.
\eeq

We now estimate the size of couplings in the $D$-dimensional effective
theory assuming that at energies $E \sim \La$, loops of all kinds are
unsuppressed.
The effective theory is perturbative for
$E \ll \La$ because of kinematic suppressions.
We can immediately write down the form of the lagrangian under this
assumption by noting that an overall factor in front of the lagrangian
acts as a loop-counting parameter, like $\hbar$ in the semiclassical
expansion.
Therefore, the lagrangian takes the form
\beq\eql{DNDA}
\scr{L}_D \sim \frac{\La^D}{\ell_D}
\hat{\scr{L}}_{\rm bulk}(\hat{\Phi}, \partial/\La)
+ \de^{D - 4}(y) \frac{\La^4}{16\pi^2}
\hat{\scr{L}}_{\rm brane}(\hat\phi, \hat\Phi, \partial/\La).
\eeq
Here, $\hat\Phi$ is a bulk field, $\hat\phi$ is a brane field,
and all couplings in the `reduced lagrangians'
$\hat{\scr{L}}_{\rm bulk}$ and $\hat{\scr{L}}_{\rm brane}$ are
order 1.

For example, comparing the form of the effective lagrangian \Eq{DNDA}
to the definition of the $D$-dimensional Planck scale
\beq
\scr{L}_D = -\sfrac{1}{2} M_D^{D - 2} \scr{r}^{(D)} + \cdots
\eeq
immediately gives
\beq\eql{Lascale}
\La \sim \ell^{1/(D - 2)} M_D.
\eeq
Numerically, $\La \sim 10 M_D$ for $5 \le D \le 11$.

The fields $\hat\Phi$ and $\hat\phi$ appearing in \Eq{DNDA} have been
taken to be dimensionless;
their kinetic terms have the form
\beq
\scr{L}_D \sim \frac{\La^{D - 2}}{\ell_D} (\partial \hat\Phi)^2
+ \de^{D - 4}(y)\frac{\La^2}{16\pi^2} (\partial\hat\phi)^2 + \cdots.
\eeq
Fields with canonical kinetic terms can be defined by
\beq
\Phi \sim \frac{\La^{(D - 2)/2}}{\sqrt{\ell_D}} \hat\Phi,
\qquad
\phi \sim \frac{\La}{4\pi} \hat\phi.
\eeq
When the lagrangian \Eq{DNDA} is expressed in terms of canonical
fields, the interactions contain nontrivial geometrical factors.

The prefactors in the lagrangian \Eq{DNDA} give rise to enhancement
factors for loops that cancel the kinematic loop suppression factors.
This is clear for diagrams involving only bulk fields or only brane
fields.
For diagrams with bulk fields coupled to brane fields, this is less
obvious, and we will discuss this point briefly.
The simplest set-up where we can understand \Eq{DNDA} is where all
$D$ dimensions are non-compact.
Since the coefficients we are estimating arise at the scale
$\La \gg 1/r$, this approximation is sufficient.
The theory may contain linear or quadratic terms in the bulk fields
localized on the wall, such as
$\De\hat{\scr{L}}_{\rm brane}
\sim \hat\Phi + \hat\Phi \hat\phi + \hat\Phi^2$.
We will treat these as perturbations for purposes of estimating
the kinematic factors.
In this case, the propagators for the bulk and brane fields are
exactly the same as in theories without branes:
\beq\bal
\avg{\Phi(X_1) \Phi(X_2)} &= \myint \frac{d^D\! P}{(2\pi)^D}\,
\frac{i}{P^2 - M^2} e^{-i P \cdot (X_1 - X_2)},
\\
\avg{\phi(x_1) \phi(x_2)} &= \myint \frac{d^4 k}{(2\pi)^4}\,
\frac{i}{k^2 - m^2} e^{-i k \cdot (x_1 - x_2)}.
\eal\eeq
(Na\"\i ve dimensional analysis suggests that $M$ and $m$ are
either vanishing or of order $\La$.)
The Feynman rules for the momentum-space correlation functions are
also the same as in theories without branes, except that in
couplings of bulk fields to the brane, the momenta perpendicular to the
brane is not conserved.
If the bulk field line coming from the brane is part of a loop,
the perpendicular components of the momenta are freely integrated over.

\begin{figure}[t] \centering %
\centerline{\epsfxsize=5.0in\epsfbox{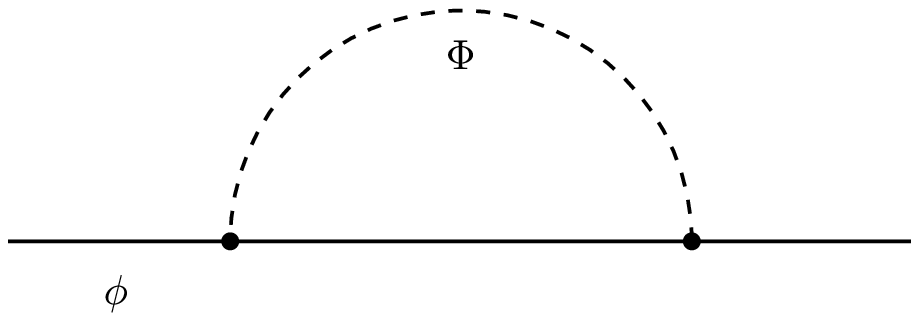}}
\smallskip
\capt{One-loop diagram involving both bulk fields $\Phi$ and
brane fields $\phi$.}
\end{figure}

For example, consider a coupling (written in terms of canonically
normalized fields) 
\beq
\De\scr{L}_D \sim \de^{D - 4}(y) \frac{\sqrt{\ell_D}}{\La^{(D - 6)/2}}
\phi^2 \Phi.
\eeq
We can then consider a `mixed' loop diagram such as the one shown in
Fig.~2.
Using the rules above, we obtain
\beq
\hbox{\rm Fig.~2} &= \left(
\frac{\sqrt{\ell_D}}{\La^{(D - 6)/2}} \right)^2
\myint \frac{d^D\! P}{(2\pi)^D}\, \frac{i}{P^2 - M^2}
\frac{i}{(P_\parallel + k)^2 - m^2}
\nonumber\\
&\sim \frac{\ell_D}{\La^{D - 6}} \frac{1}{\ell_D}
\left[ \La^{D - 4} + \La^{D - 6} k^2 + \cdots \right],
\eeq
where $k$ is a 4-momentum on the brane and $P_\parallel$ is the
projection of the bulk momentum $P$ in the direction parallel to the
3-brane.
The kinematic suppression from the loop cancels the enhancement
factor from the coupling, and the result is the same order of
magnitude as the tree-level term.

\subsection{Estimates of Soft SUSY Breaking}
We now estimate the size of the various parameters in this theory.
We first consider the size of the extra dimensions.
The extra dimensions must be large enough to suppress FCNC's, but
small enough so that the standard-model gauge bosons (which
propagate in the bulk) have 4-dimensional gauge couplings of
order 1.
Using the results of the previous Subsection, the maximum value
$D$-dimensional gauge coupling is
\beq
g_{D,{\rm max}}^2 \sim \frac{\ell_D}{\La^{D - 4}}.
\eeq
Since $g_4^2 \sim g_D^2 / V_{D - 4} \sim 1$, this gives a
maximum value for the volume of the extra dimensions.
For a symmetric toroidal compactification the volume of
the compact space is $V_{D - 4} = L^{D - 4}$, where $L$ is the length
of the sides, and we obtain
\beq
L_{\rm max} \sim \frac{\ell_D^{1/(D - 4)}}{\La}.
\eeq
For a spherical compactification with radius $r$, we obtain
\beq
r_{\rm max} \sim \left(
\frac{\ell_D}{\Om_{D - 3}} \right)^{1/(D - 4)}
\frac{1}{\La}.
\eeq
Numerically, $L_{\rm max}, r_{\rm max} \gsim 10/\La$ for all
$5 \le D \le 11$.
This is sufficiently large to suppress FCNC's, since these
are exponentially suppressed by Yukawa factors.
See Table 1.
Note that for $D = 5$ or 6, we do not need the $D$-dimensional gauge
coupling to be strongly coupled at the scale $\La$, but strong
coupling is required for larger values of $D$.


\begin{table}[t] 
\centering
\begin{tabular}{c|cc|cc}
$D$ & $\La L_{\rm max}$ & $e^{-\La L_{\rm max}/2}$
& $\La r_{\rm max}$ & $e^{-\La r_{\rm max}}$ \\
\hline
5 & 740 & $3 \times 10^{-162}$ & $118$ & $4 \times 10^{-52}$ \\
6 & 63 & $2 \times 10^{-14}$ & $18$ & $2 \times 10^{-8}$ \\
7 & 29 & $6 \times 10^{-7}$ & $11$ & $3 \times 10^{-5}$ \\
8 & 20 & $5 \times 10^{-5}$ & $8.7$ & $2 \times 10^{-4}$ \\
9 & 16 & $3 \times 10^{-4}$ & $8.0$ & $3 \times 10^{-4}$ \\
10 & 14 & $9 \times 10^{-4}$ & $7.8$ & $4 \times 10^{-4}$ \\
11 & 13 & $1 \times 10^{-3}$ & $7.8$ & $4 \times 10^{-4}$ \\
\end{tabular}
\capt{Estimates for the toroidal compactification length $L$
and spherical compactification radius $r$, as well as the
exponential suppression factor for massive propagation
between two branes of maximal separation.}
\end{table}

We now estimate the size of the soft SUSY breaking terms.
We first consider the possibility that the theory is weakly coupled
at the $D$-dimensional Planck scale $M_D$, and that all couplings
are order 1 in units of $M_D$.
Actually, the bulk gauge coupling must be somewhat larger than
this in order that $g_4 \sim 1$, but \eg~for $D = 5$ we require
only $g_5^2 \gsim 10/ M_D$ in order to obtain $L M_D \gsim 10$.
In this case, it is plausible that the VEV that breaks $U(1)_X$ is
$v \sim M_D$, and that the contact terms \Eqs{Dgaugino} and \eq{Dmu}
are also order 1 in units of $M_D$.
In this case, all soft masses are of order
\beq
m_{\rm soft} \sim \frac{\avg{F_\Si}}{M_D} \sim m_{3/2}
\left( V_{D - 4} M_D^{D - 4} \right)^{1/2},
\eeq
where the 4-dimensional Planck scale $M_4$ is given by
\beq\eql{M4}
M_4^2 = V_{D - 4} M_D^{D - 2}.
\eeq
If the compact dimensions are very large, the gravitino is the LSP,
although this is easily avoided for $D = 5$ or $6$.

We now consider the alternative that all couplings in the theory
are strong at the fundamental scale $\La$.
We first estimate the value of $\La$ in this scenario.
In this case, the gauge couplings are as large as possible so we
take $r \sim r_{\rm max}$.
Combining this with the estimate of $\La$ given in \Eq{Lascale} and the
formula for the 4-dimensional Planck scale \Eq{M4} we obtain the simple
result
\beq\eql{Laresult}
\La \sim M_4.
\eeq

We now estimate the standard-model scalar masses, gaugino masses, and
$\mu$ and $B\mu$ terms.
The gaugino mass and $\mu$ and $B\mu$ terms are estimated from the
coefficients of the higher-dimension operators \Eqs{Dgaugino} and
\eq{Dmu} connecting the bulk gauge and Higgs fields to the fields
propagating on the 3-branes.
Using the estimates for strongly coupled theories given in the
previous Subsection, we find that the effective 4-dimensional theory
(written in terms of canonically normalized fields) contains the terms
\beq\bal
\scr{L}_{4} \sim & \myint d^4\th \left[
\frac{1}{4\pi\La} \Si^\dagger
(H_u H_d
+ H_u^\dagger H^{\vphantom\dagger}_u
+ H_d^\dagger H^{\vphantom\dagger}_d)
+ \frac{1}{\La^2} \Si^\dagger \Si H_u H_d \right]
\\
&\quad
+\, \myint d^2\th\, \frac{1}{4\pi\La} \Si W^\al W_\al + \hc
\eal\eeq
Here the relation $g_4^2 \sim 1$ has been used to
eliminate the dependence on the $D$-dimensional loop counting parameter
$\ell_D$.
We therefore find
\beq
m_\la \sim \mu
\sim \frac{1}{4\pi} m_{3/2},
\qquad
m_H^2 \sim B\mu 
\sim m_{3/2}^2.
\eeq
It is important that these soft masses are
larger than the anomaly-mediated contributions.
For example, anomaly mediation gives a contribution to gaugino
masses $\De m_\la / m_\la \sim 1 / (4\pi)$.

For the scalar masses, we use the fact that the natural size for the
VEV of a dimensionless wall field in a strongly-coupled theory is
$\avg{\hat\phi} \sim 1$ \cite{SUSYNDA}.
This gives a $U(1)_X$ breaking VEV of order
$v \sim \La / (4\pi)$,
and squark and slepton masses are
\beq\eql{scalar}
m^2_{\tilde{Q},\tilde{\ell}}
\sim 16\pi^2 X_Q X_\Si m_{3/2}^2,
\eeq
where $X$ is the $U(1)_X$ gauge charge.
These estimates give
$m_{\tilde{Q}} / m_\la \sim 16\pi^2 (X_Q X_\Si)^{1/2}$.
If we take this at face value, we must choose the $U(1)_X$
charges of $Q$ and $\Si$ to be small in order to avoid unrealistically
large squark masses.%
\footnote{Recall that the $U(1)_X$ charges are normalized so that
the field that obtains a VEV has $X = +1$.}
However, we should allow uncertainties in the estimates of
strong-interaction quantities at the level of an order of magnitude,
in which case we can easily obtain scalar and gaugino masses
of the same order with only moderately small $U(1)_X$ charges.
In addition, there is the possibility that there are other moderately
large factors that modify this result.
For example, large-$N$ counting in the sector that breaks the $U(1)_X$
gives $v \sim \sqrt{N} \La / (4\pi)$, which reduces $m_{\tilde{Q}}^2$ by a
factor of $N$.
We expect realistic models will have a large number of degrees of freedom
(\eg~$N \gsim 10$) that will affect the estimates above in other sectors
as well.

The above estimate for the scalar masses uses the results of the
tree-level calculation of Section 2.
It is important to know whether these results are qualitatively
reliable in the strongly-coupled case we are considering.
The key point is that the squark masses arise from a non-local
effect in the $D$-dimensional theory, and are therefore insensitive
to the short-distance physics.
In particular, the leading contribution to the scalar masses arises
from the exchange of the lightest $U(1)_X$ KK mode, which has mass
of order $1/r \ll \La$.
As long as the strong dynamics gives rise to these light states
with the symmetry breaking patter assumed, we expect the
estimates above to be valid at the order-of-magnitude level.

Another attractive possibility is that the couplings of fields on
the 3-branes are perturbative (dimensionless couplings of order 1),
while the bulk fields are strongly coupled.
In this case, the estimates for the gaugino masses and $\mu$ and
$B\mu$ terms are the same as in the strongly-coupled scenario discussed
above.
However, in this case, the natural size for the VEV that breaks
$U(1)_X$ is $v \sim \La$, which gives
$m_{\tilde{Q}} / m_\la \sim 4\pi (X_Q X_\Si)^{1/2}$.

Of course, the set-up we have described does not explain all small
parameters in the low-energy effective field theory.
For example the estimates above tell us that the Yukawa couplings
are order 1 (even in the strongly-coupled case).
This is a good starting point for a theory of flavor, since it can
explain why the top Yukawa coupling is perturbative but order 1,
but clearly additional structure is needed to explain why the
other Yukawa couplings are suppressed.
There are also other small numbers (\eg~in cosmology) that are not
explained in this scenario as elaborated so far.
It would be interesting to see if there are higher-dimensional mechanisms
that can explain these hierarchies and small numbers in our scenario,
perhaps analogous to those considered in the context of millimeter-sized
extra dimensions \cite{mm}.

\subsection{Phenomenology}
We now comment briefly on the phenomenology of these models.

The first important point is that the scalar masses generated
by $U(1)_X$ exchange are naturally flavor-diagonal if $U(1)_X$ is
broken in the hidden sector and if $U(1)_X$ commutes with flavor
symmetries.
This is perhaps the most attractive feature of the present class
of models.

If the Yukawa couplings arise from $U(1)_X$ invariant effects,
the fact that the down-type quarks and the leptons get masses from the
same Higgs field implies the scalar mass relation
\beq
m^2_{\tilde{L}L} + m^2_{\tilde{e}R}
= m^2_{\tilde{Q}L} + m^2_{\tilde{d}R},
\eeq
up to small radiative corrections.%
\footnote{The radiative corrections may be significant for large
$\tan\be$.}
This is the same as a $SU(5)$ GUT relation, but in the present models
it may hold even in the absence of grand unification (\eg in string
unification).

Another general feature of the present models is that
$A$ terms can naturally be small.
In conventional hidden sector models these are generated by operators
of the form
\beq
\De\scr{L}_{\rm eff} \sim \myint d^2\th\,
\frac{1}{M} \Si QQ H_{u,d} + \hc,
\eeq
but in the present models these are exponentially suppressed because
the hidden and visible sectors are separated.
$A$ terms can arise from the operator
$\int\! d^4\th (\Si^\dagger X \Si) (Q^\dagger X Q)$
in models where $\avg{\Si} \ne 0$ in addition to
$\avg{F_\Si} \ne 0$.
In such models, we expect $\avg{\Si} \ll M_D$, so this contribution is
also suppressed.
There is an unavoidable contribution to the $A$ terms from
anomaly mediation \cite{GLMR}, but it is suppressed both by loop
factors and Yukawa couplings, and is therefore negligible for
most purposes.

Another possibility is that the $\mu$ and $B\mu$ terms are
generated by the VEV of a singlet $S$ in the visible sector.
Models of this type require the following superpotential terms
in the effective 4-dimensional theory:
\beq
W_{\rm obs} = \la S H_u H_d + \frac{\ka}{3} S^3
+ \hbox{\rm Yukawa\ couplings}
\eeq
The terms involving $S$ are not $U(1)_X$ invariant, but they can arise with
couplings of order 1 if the Higgs propagates in the bulk and $S$
propagates either in the bulk or on the hidden-sector brane.
In that case, these terms can arise from higher-dimension brane
superpotential terms involving the field that breaks $U(1)_X$.

One very attractive feature of this model is that the absence of
large $A$ terms implies that it solves the SUSY CP problem.
In this model, the only terms in the effective lagrangian with
possible CP-violating phases are the visible sector superpotential
couplings and the hidden-sector superpotential term that gives rise
to gaugino masses
\beq
W_{\rm hid} = \frac{c}{M_D} \Si W^\al W_\al.
\eeq
(Here we assume that the theory is embedded in a GUT so that there is
only one independent gaugino mass.)
The phases in $\la$, $\ka$, and $c$ can be rotated away as follows.
A $U(1)_R$ rotation (where all matter fields have $r = +\sfrac 23$)
can be used to make the gaugino mass real.
Then $S$ can be rephased to make $\ka$ real.
Finally, we use a $U(1)_{\rm PQ}$ rotation to make $\la$ real,
where the PQ charge of all quark and lepton fields is $-\sfrac 12$,
$H_{u,d}$ have charge $+1$, and $S$ has vanishing PQ charge.
Note that these transformations will not eliminate phases in the $A$
terms in general, but we have seen that it is natural for the $A$
terms to arise only from anomaly mediation.
This means that they are loop suppressed, and also their phases come
from the phases in the superpotential couplings.

If the $\mu$ problem is solved by the Giudice-Masiero mechanism,
there are uncontrolled phases in the gaugino mass and $\mu$ and
$B\mu$ terms that cannot be eliminated by field redefinitions.
These models therefore have a SUSY CP problem identical to conventional
hidden sector models.

In many ways the phenomenology of these models is very similar to
conventional hidden sector models:
scalar and gaugino masses are of order $m_{3/2}$, and in the context
of GUT models, scalar masses in the same GUT multiplet and gaugino
masses unify at the GUT scale.
In fact, since the scalar mass-squared terms are controlled by $U(1)_X$ gauge
charges, it is natural for them to be equal (or have simple rational ratios)
at the $U(1)_X$ breaking scale even in the absence of grand unification
(\eg in string unification).
This may also occur in `anomalous $U(1)$' models \cite{anomuone}.
This gives the possibility of a rather distinctive signature, namely
that scalar masses unify while gaugino masses do not.

FCNC's are exponentially suppressed in this model, and are therefore
unobservably small unless the value radius accidentally puts them
near the experimental limits.

Finally, we mention that the $D - 4$ `extra' polarizations of the bulk
$U(1)_X$ gauge field ($A^I$, $I = 4, \ldots, D-1$) do not get mass from the
Higgs mechanism on a 3-brane, and therefore may be light.
The same may be true for other components of the supersymmetric gauge
multiplet (\eg~gauginos).
In the specific model constructed in Section 3.1 these states obtain
masses of order the compactification scale by the orbifold projection.
However, it may be interesting to consider other scenarios where
these fields are light.
By gauge invariance, the fields $A^I$ can have non-derivative couplings
only to charged bulk fields.
If there are no charged bulk fields, these fields are derivatively
coupled (through the $U(1)_X$ gauge field strength) with higher-dimension
operators suppressed by powers of $M_D$.
Such fields are not visible in terrestrial experiments and
will not be in equilibrium in the early universe provided that
the inflationary reheat temperature is smaller than $M_D$.
If the Higgs fields propagate in the bulk and are charged under
$U(1)_X$, there is an important coupling from the Higgs kinetic term
$D^M H^\dagger D_M H = H^\dagger H A^I A_I + \cdots$.
In fact, this interaction gives the fields $A^I$ a (positive)
weak-scale mass.
This scenario is very constrained, especially since we expect that
supersymmetric partners of the $A^I$ will also be light.
Analogous remarks are expected to hold for supersymmetric partners
of the $A^I$ fields.
Since these possibilities are highly model-dependent, we will not
analyze them further here.

\section{Conclusions}
We have argued that a bulk $U(1)_X$ gauge field broken on the
hidden-sector 3-brane is an attractive candidate for the messenger of
supersymmetry breaking.
This scenario automatically suppresses flavor-changing neutral currents
independently of the flavor structure at the fundamental Planck scale,
while at the same time naturally giving positive scalar
mass-squared terms.
Gaugino masses are naturally generated if the standard-model gauge
fields propagate in the bulk.
The $\mu$ problem can be solved either by the Giudice-Masiero mechanism
if the Higgs fields propagate in the bulk, or by the VEV of a field
in the visible sector.
In the latter case, the SUSY CP problem is automatically solved because
of the absence of large $A$ terms.

In these models, the scales of the supersymmetry breaking parameters can be
naturally related if we assume that all microscopic parameters are order 1
at the fundamental Planck scale.
Another natural possibility is that all interactions in the theory
are strongly-coupled at a fundamental scale near the $D$-dimensional
Planck scale.
In this scenario, the only large hierarchy is the `large' size of the
extra dimensions, which need only be an order of magnitude compared to
the fundamental scale.
This is attractive from the point of view of string theory, where there
are severe difficulties in formulating realistic theories at weak
coupling.


\section*{Acknowledgments}
We thank R. Rattazzi for pointing out a mistake in an earlier
version of this paper.
M.A.L. thanks Stanford University, Lawrence Berkeley National
Laboratory, and the Institute for Theoretical Physics at
Santa Barbara for hospitality during the course of this work.
This work was supported by the National Science Foundation under
grant PHY-98-02551, and by the Alfred P. Sloan Foundation.


\end{document}